\documentclass[12pt,a4paper]{article}
\usepackage[T2A]{fontenc}
\usepackage[cp1251]{inputenc}
\usepackage{amsmath,amssymb,amsthm,amsfonts,amscd}

 \theoremstyle{plain}
\newtheorem{theorem}{Theorem}
\newtheorem*{main*}{Main Theorem}
\newtheorem{denotation}{Denotation}
\newtheorem*{theorem*}{Theorem}

\newtheorem{proposition}[theorem]{Proposition}

\theoremstyle{definition}
\newtheorem{definition}[theorem]{Definition}
\newtheorem{remark}[theorem]{Remark}
\newtheorem*{remark*}{Remark}
\newtheorem*{example*}{Example}
\newtheorem{example}[theorem]{Example}

\def\Z{{\Bbb Z}}
\def\R{{\Bbb R}}

\def\A{{\bf A}}
\def\D{{\bf D}}

\def\k{{\bf k}}
\def\i{{\bf i}}

\def\rot{{\operatorname{rot}}}
\def\R{{\Bbb R}}
\def\d{{\bf d}}
\def\B{{\bf B}}

\def\x{{\bf x}}

\begin{document}
\sloppy

\title{Dispersion of the Arnold's asymptotic ergodic Hopf invariant and a formula for its calculation} 
\author{
P. M. Akhmet'ev\footnote{National Research University "Higher School of Economics", Moscow, Moscow, Russia; IZMIRAN, Troitsk, Moscow region, Russia; pmakhmet@mail.ru} 
\- and  	I. V. Vyugin \footnote{National Research University "Higher School of Economics", Moscow, Russia;	
Institute for Information Transmission Problems of the Russian Academy of Sciences (Kharkevich Institute), Moscow, Russia}}
\maketitle

\begin{abstract}
The paper contains an example to describe magnetic fields in a conductive medium. The authors assume that  new applications for turbulent magnetic fields in the case the magnetic field is not left- and right- polarized are possible.
\end{abstract}

\section{Introduction}

The paper contains an example, which describes magnetic fields in a conductive medium.
The basic equations and the problem can be found in \cite{Arn-Kh}. A new application assumes that the Fourier spectra of magnetic fields are random. This assumption is analogous to the hydrodynamic turbulence introduced by A.N.Kolmogorov, see \cite{Fr}. The situation with magnetohydrodynamic turbulence is more complicated and Arnold's asymptotic ergodic Hopf invariant is very important. The asymptotic Hopf invariant is called the magnetic helicity, this magnetic helicity is denoted by $\chi_{\B}$. The definition of the magnetic helicity is in \cite{Arn-Kh}, we recall it in the formula $(\ref{helicity})$.  
Basic constructions for magnetohydrodynamic turbulence  are in   \cite{Fr-S}. 
Example \ref{Ex1}, Section \ref{Sect4} illustrates the importance of magnetic helicity.

Magnetic lines have complicated geometry. A distribution function of asymptotic linking numbers is said to be  random. 
The approach by V.I.Arnold shows that the helicity is the mean value of the distribution of asymptotic linking numbers of magnetic lines. Dispersions of the distribution are interesting. The dispersions of asymptotic numbers of magnetic lines are called the quadratic helicities. In the paper we investigate one of the two dispersions, denoted by $\chi^{[2]}_{\B}$, or, by $\chi^{[2]}$.  
A new Example \ref{Ex2} is analogous to Example \ref{Ex1}, this example illustrates the importance of the quadratic magnetic helicity for magnetohydrodynamic turbulence. 

The authors are grateful to  A.V.Parusnikova for discussions and to the Reviewer for remarks.  Work of P.M.Akhmet'ev was supported in part by RFBR GFEN 17-52-53203.
Work of I.V.Vyugin was supported by the Russian Science Foundation Grant  18-41-05003.

\section{The Arnold's asymptotic Hopf invariant and its random distribution}

Let us recall definition of asymptotic linking number of a pair of trajectories as in \cite{Arn-Kh}, Ch.III.
Let $\B$ a divergent-free (magnetic) field in 3D domain $\Omega \subset \R^3$. We assume that $\B$ is tangent to the boundary $\partial \Omega$ and has no zeros.  Denote by $g^t: \Omega \to \Omega$ the phase flow of $\B$. 
Take two points $x_1,x_2 \in \Omega$.

\begin{definition}\label{def1}
The asymptotic linking number of the pair of trajectories  $g^t(x_1)$, $g^t(x_2)$ is denoted by the limit
\begin{eqnarray}\label{lim}
\lambda_{\B}(x_1,x_2) = \lim_{T \to + \infty} \frac{lk_{\B}(x_1,x_2;T)}{T^2},
\end{eqnarray}
where  
\begin{equation}\label{Gauss}
\begin{array}{cc}
lk_{\B}(x_1,x_2;T) = \frac{1}{4\pi} \int_0^T \int_0^T \frac{(\dot{\gamma}_1,\dot{\gamma}_2,\gamma_1-\gamma_2)}{\vert\vert \gamma_2 - \gamma_1\vert\vert^3} \d t_1 \d t_2 =\\
 \int_0^T \int_0^T G(x_1(t_1),x_2(t_2)) \d t_1 \d t_2, \quad \dot{\gamma}_i(t_i) = \B(g^{t_i}(x_i)), \quad i=1,2,
\end{array}
\end{equation}
\end{definition}
is the linking number  of the two segments $\gamma_1 = g^{t_1}(x_1); t_1 \in [0,T]$, $\gamma_2 = g^{t_2}(x_2); t_2 \in [0,T]$.
The integral in the formula is called  the Gauss integral. In the case the trajectories $\gamma_1$, $\gamma_2$ parametrizes closed circles of the unite length, the limit $(\ref{Gauss})$ coincides with the linking number of this two circles, which is a topological invariant.

Let us denote  the point $g^{t_i}(x_i)$ by $x_i(t_i), \quad i=1,2$.
In the right-hand side of the formula $(\ref{Gauss})$ by 
\begin{eqnarray}\label{G}
G = \frac{1}{4\pi} \frac{(\dot{\gamma}_1,\dot{\gamma}_2,\gamma_1-\gamma_2)}{\vert\vert \gamma_2 - \gamma_1\vert\vert^3} \end{eqnarray} 
is denoted the kernel of the Gauss integral. The denotation $G(x_1,x_2)=G(\B(x_1),\A(x_2;x_1))$, where 
$$\A(x_2;x_1)= \frac{1}{4\pi} \frac{\B(x_2) \times (x_1-x_2)}{\vert\vert x_1 - x_2 \vert\vert^3} $$ 
is the Biot-Savart potential, it will be used below in the formula $(\ref{tensor})$.

Let us consider the domain of such points $(x_1,x_2)$ in $\Omega \times \Omega$ that $\lambda_{\B}(x_1,x_2)$ is well-defined. 
If the trajectory issued from $x_1$ contains the point $x_2$, the function $\lambda_{\B}(x_1,x_2)$ is not defined. The domain of $\lambda_{\B}$ is a measurable subset in $\Omega \times \Omega$.  The ergodic theorem implies that the function $\lambda_{\B}: \Omega \times \Omega \to \R$ is well defined almost everywhere,
and belongs to the space $L^1$. This follows from the fact that the function $(\ref{Gauss})$ belongs to $L^1$.

A dimension of $\lambda_{\B}$ is $G^2 \cdot cm^{-2}$. 
This means that the transformation $\B \mapsto l\B$, $\x \mapsto m\x$, $\x \in \R^3$ determines the transformation 
$\lambda_{\B} \mapsto l^2 m^{-2} \lambda_{\B}$ of the asymptotic linking number. In the CGS system magnetic field is measured into Gaussian units $G$. 
The average self-linking number 
\begin{eqnarray}\label{helicity}
\chi_{\B} = \int \int \lambda_{\B}(x_1,x_2) \d \Omega \d \Omega
\end{eqnarray} 
of a magnetic field $\B$ in $\Omega$ is called the asymptotic Hopf invariant, or, the helicity. 
The helicity is a lower bound of the magnetic energy by the Arnold inequality \cite{Arn-Kh} Section III, Theorem 1.4.
For a divergence-free vector field (a magnetic field) $\B$,
\begin{eqnarray}\label{Arnoldineq}
\int_{\Omega} (\B(x),\B(x)) \d V \ge C \vert \chi_{\B} \vert, 
\end{eqnarray}
where $C$ is a positive constant dependent on the shape and size of the compact domain $\Omega$
with a magnetic field. In the right hand side of the formula $(\ref{Arnoldineq})$ we have the invariant of volume-preserved transformation of the domain. In the left hand side of the formula we have the magnetic energy. The inequality proves that the absolute value of the magnetic helicity  $\chi_{\B}$ determines a lower boundary of  the magnetic energy. Example \ref{Ex1}
for magnetohydrodynamic turbulence is an analogous one.

In \cite{Arn} (the bottom remark in the subsection Example 5.2) is mentioned that the function $m(\lambda_0)$, defined as the measure of the set $\{(x_1,x_2) \in \Omega \times \Omega \vert \lambda_{\B}(x_1,x_2)<\lambda_0\}$, is the much stronger invariant of volume-preserved transformations than the helicity. A lower bound of the magnetic energy, which
is calculated using this distribution function, is more
 sharp than the bound, which is calculated using the magnetic helicity $\chi_{\B}$ in the Arnold inequality. 

The function $m(\lambda_0)$ is a distribution function of asymptotic linking numbers.   Using ergodic theorem 
for the function $G(x_1,x_2)$  with respect to the flow $g^{t_1} \times g^{t_2}$ in $\Omega \times \Omega$ one may say only, that $m(\lambda_0)$ admits a mean value: the helicity. But, what to do  if dispersion of this distribution is well-defined? We will get an affirmative answer to this question.

\section{Quadratic helicity; a local formula}

Formally, the dispersion of the asymptotic self-linking number $\lambda_{\B}(x_1,x_2)$ is defined by the integral:

\begin{eqnarray}\label{dispersion}
\iint (\lambda_{\B}(x_1,x_2) - \frac{\chi_{\B}}{Vol(\Omega)})^2 \quad \d \Omega \d \Omega,
\end{eqnarray} 
where $Vol(\Omega)$ is the volume of the domain $\Omega$. In this formula $\frac{\chi_{\B}}{Vol(\Omega)}$
is the average value of the linking numbers of magnetic line in $\Omega$. Obviously, $(\ref{dispersion})$ is equivalent
to the integral
\begin{eqnarray}\label{dispersion1}
\iint \lambda^2_{\B}(x_1,x_2)\quad \d \Omega \d \Omega - \chi^2_{\B}.
\end{eqnarray}
The first term in $(\ref{dispersion1})$ is called (a half of) the quadratic helicity:
\begin{eqnarray}\label{dispersion2}
\chi^{[2]}_{\B} = 2\iint \lambda^2_{\B}(x_1,x_2)\quad \d \Omega \d \Omega. 
\end{eqnarray}
In \cite{Akh} it is proved that $\chi^{[2]}$ (a dimension of $\chi^{[2]}$ is $G^4 cm^2$) is well-defined. Also, in this paper an inequality between 
$\chi_{\B}$ and $\chi^{[2]}_{\B}$ is proved. 
The goal of this section is a generalization a local formula from \cite{A-C-Sm} for the quadratic helicity $\chi^{(2)}$ (see Section 4 for
a brief definition and \cite{Akh} for definition) for $\chi^{[2]}$.

Let us recall definition of $\delta_{\B}^{[2]}$, which is called (a component of) the magnetic correlation tensor: 
\begin{eqnarray}\label{tensor}
\delta^{[2]}_{\B}(x_1,x_2) = G^2(x_1,x_2) = (\B(x_1),\A(x_2;x_1))^2 = (\A(x_1;x_2),\B(x_2))^2,
\end{eqnarray}
where $\A(x;y)$ is the Biot-Savart potential as in the formula $(\ref{Gauss})$, see \cite{Arn-Kh} Ch 3 Paragraph 4;
$G^2(x_1,x_2)$ is the square of the Gaussian kernel. Let us recall that
$(\B(x_1),\A(x_2;x_1))$ is the kernel  $G(x_1,x_2)$ of the Gauss integral $(\ref{G})$. 
 The inequality
\begin{equation}\label{noneq}
 \chi^{[2]}_{\B} \le \iint \delta^{[2]}_{\B} \d\Omega \d \Omega
\end{equation} 
is proved in $\cite{Akh}$. This proof follows from the fact that the function $\delta^{[2]}_{\B}(x_1,x_2)$ is integrable over $\Omega \times \Omega$. 

The correlation tensor 
$$\delta^{[2]}_{\B_1,\B_2} = (\B_1(x_1),\A_2(x_2;x_1))^2=(\A_1(x_1;x_2),\B_2(x_2))^2$$
 for a pair  $\B_1$, $\B_2$ of magnetic fields is not  
integrable over $\Omega \times \Omega$, because the asymptotic of $\delta^{[2]}_{\B_1,\B_2}(x_1,x_2) \simeq \vert \vert x_1 - x_2 \vert \vert^{-4} $, when $x_1 \to x_2$. 
One may define the symmetrization  $\delta^{[2];sim}_{\B_1,\B_2}(x_1,x_2)$ by the formula:
\begin{eqnarray}\label{sim}
\delta^{[2];sim}_{\B_1,\B_2}(x_1,x_2) = \frac{1}{2}[G_{\B_1+\B_2}(x_1,x_2)-G_{\B_1}(x_1,x_2) - G_{\B_2}(x_1,x_2)]^2
\end{eqnarray} 
where $G_{\B_1+\B_2}$, $G_{\B_1}$, $G_{\B_2}$ are the kernels of the Gauss integral for corresponding vector-fields.
We will use this denotation not only for magnetic fields $\B_1$, $\B_2$, which are divergent-free, but in general cases.  
In the case $\B_1=\B_2=\B$, we get $\delta^{[2];sim}_{\B,\B}(x_1,x_2)=\delta^{[2]}_{\B,\B}(x_1,x_2)$.

%The correlation tensor $$\delta^{[2]}_{\B_1,\B_2,\B'_1,\B'_2} = (\B_1(x_1),\A_2(x_2;x_1))(\B'_1(x_1),\A'_2(x_2;x_1))$$ %for a quadruple  $\B_1$, $\B_2$, $\B'_1$, $\B'_2$ of magnetic fields is not  
%integrable over $\Omega \times \Omega$.
% because the asymptotic of $\delta^{[2]}_{\B_1,\B_2}(x_1,x_2) \simeq \vert \vert x_1 - x_2 \vert \vert^{-4} $, when %$x_1 \to x_2$. 
%One may define the symmetrization  $\delta^{[2];sim}_{\B_1,\B_2,\B'_1,\B'_2}(x_1,x_2)$, which is integrable, by the %formula:
%\begin{eqnarray}\label{sim}
%\delta^{[2];sim}_{\B_1,\B_2,\B'_1,\B'_2}(x_1,x_2) = \frac{1}{2}[\delta^{[2];sim}_{\B_1+\B'_1,\B_2+\B'_2}(x_1,x_2)- %\delta^{[2];sim}_{\B_1,\B_2}(x_1,x_2)-\delta^{[2];sim}_{\B'_1,\B'_2}(x_1,x_2)]. 
%\end{eqnarray} 
%In the case $\B_1=\B_2=\B'_1=\B'_2$, we get $\delta^{[2];sim}_{\B,\B,\B,\B}(x_1,x_2)=\delta^{[2]}_{\B,\B}(x_1,x_2)$.

Let us denote
\begin{equation}
\delta^{[2]}(\B_1,\B_2) = \iint \delta^{[2];sim}_{\B_1,\B_2}(x_1,x_2) \d \Omega \d \Omega.
\end{equation}

\begin{denotation}\label{denot}
Denote the vector-field 
$\vert\vert x_1 - x_2 \vert\vert^3 G(x_1,x_2)$ 
by $\tilde{G}(x_1,x_2)$, the function $\vert\vert x_1 - x_2 \vert\vert^{-3}$ by $\Psi(x_1,x_2)$.
Denote by $\nabla_{\B}(\D)$ the derivative (of components) of the vector $\D$ along the vector $\B$.
Using this we introduces the following denotations:
$$\B^{(i)}_1(x_1,x_2))=\Psi^{-1}(x_1,x_2)\nabla_{\B}^i [\Psi(x_1,x_2)\B(x_1)],$$ $$\B^{(j)}_2(x_1,x_2))=\Psi^{-1}(x_1,x_2)\nabla_{\B}^j [\Psi(x_1,x_2)\B(x_2)],$$
$$\nabla^{i+1}_{\B}[\Psi(x_1,x_2)\B(x_1)] = \nabla_{\B}(\nabla^i_{\B}[\Psi(x_1,x_2)\B(x_1)],$$
$$\nabla^{j+1}_{\B}[\Psi(x_1,x_2)\B(x_2)] = \nabla_{\B}(\nabla^j_{\B}[\Psi(x_1,x_2)\B(x_2)].$$
\end{denotation}

\begin{theorem}\label{12}
With an assumption  that 

$\odot$ $(1)$ $\B$ is  smooth in $\Omega$ everywhere, except points on the boundary $\partial \Omega$ of the domain;

$\odot$ $(2)$ the function $G(x_1,x_2), \quad x_1 \in l_1, x_2 \in l_2$  in $\Omega \times \Omega \setminus diag$ (the kernel of the Gauss integral)
contains a  Fourier spectrum with wave numbers $k$ in a finite interval $\{0\} \cup [\Delta';\Delta]$, $0 < \Delta' << \Delta < + \infty$; 

the following equation is satisfied: 

\begin{equation}\label{11}
\frac{1}{2} \chi^{[2]}_{\B} =  \lim_{a \to +\infty} \sum_{s=0}^{\infty} (-a)^{s} \sum_{i,j;i+j=s} \frac{1}{i!j!} 
\delta^{[2]}(\B^{(i)}_1,\B^{(j)}_2), 
\end{equation} 
\end{theorem}
\begin{remark}
Obviously, the main term of the formula $(\ref{11})$ for $i=0$, $j=0$ is given by
$\iint G^2(x_1,x_2) \d \Omega \d \Omega$. In the formula $(\ref{11})$ the parameter $a$ has the dimension $G^{-1}cm$, with this assumption all terms in the formula have the dimension $G^4 cm^2$. 
\end{remark}

\begin{proposition}\label{rem}
Terms $\delta^{[2]}(\B^{(i)}_1,\B^{(j)}_2)$ in the right-hand side of the formula $(\ref{11})$ belong to the space $L^2(\Omega \times \Omega)$.
\end{proposition}

\subsubsection*{Proof of Proposition \ref{rem}}

When $x_1 \mapsto x_2$ the kernel of the integral $(\ref{Gauss})$ becomes singular, but the correlation tensor $(\ref{11})$ is absolutely integrable. After we take derivatives, as in
$(\ref{11})$, the integrability of terms is not obvious. Let us prove that the symmetrization of terms keeps the integrability. Take the expression $\nabla_{\B}^i [\Psi(x_1,x_2)\B(x_1)]$, for $i=1$. When $x_1 \mapsto x_2$, the function $\nabla_{\B}[\vert \vert x_1-x_2 \vert \vert^{-3}\B(x_1)]$ has the asymptotic $\vert\vert x_1 - x_2 \vert\vert^{-3}$. As the result,  the integral $\delta^{[2]}(\B^{(1)}_1,\B_2)$ converges before symmetrization. To pass to the next step, we may take the symmetrization of $\nabla_{\B}[\vert \vert x_1 -x_2 \vert \vert^{-3}\B(x_1)]$,
which has the asymptotic $\vert\vert x_1 - x_2 \vert\vert^{-2}$. This gives the induction for estimations of
 terms in $(\ref{11})$. \qed

\subsubsection*{Proof of Proposition \ref{12}}
Take an ordered marked pair of magnetic lines $l_1,l_2$. Take the natural measure $\d t_1 \d t_2$ on $l_1 \times l_2$, where $t_1, t_2$ are magnetic parameters, see Definition \ref{def1}.  Then the bottom term $i=0, j=0$ in $(\ref{11})$, restricted to the standard $[0,T] \times [0,T]$-segments in the Cartesian product $l_1 \times l_2$,  coincides with the 
the integral $(\ref{Gauss})$.  

The integral $(\ref{11})$ is a family of asymptotic integrals over pairs of magnetic lines. Take
the Cartesian product $\Omega \times \Omega \times [0,T]^2$ and define a small parameter $\delta$ and a big parameter $T$ as following. Consider a subspace $[\Omega \times \Omega \times [0,T]^2]_{\delta,T} \subset \Omega \times \Omega \times [0,+\infty]^2$, which consists of all pairs of magnetic $T$-lines with $\delta$-disjoin in $\Omega \times \Omega$.
Take the limit $\delta \to +0$, $T \to +\infty$. 
The formula $(\ref{dispersion2})$ is  an asymptotic integral over $\Omega \times \Omega \times [0,+\infty]^2$
of the kernel $G(x_1,x_2)$, which is extended to this Cartesian product as $G(x_1(t_1),x_2(t_2))$.
By the ergodic theorem the formula $(\ref{dispersion2})$
for a subdamain $[\Omega \times \Omega \times [0,T]^2]_{\delta,T}$ tends to $\chi^{[2]}$ in the limit.

Consider the Fourier base $\aleph$ in $\Omega \times \Omega \setminus diag$, as in the Condition $\odot$ (2). This base is extended
to  the Cartesian product $[(\Omega \times \Omega \setminus diag) \times [0,T]^2]$. 
Restrict this base to the subspace $[\Omega \times \Omega \times [0,T]^2]_{\delta,T} \subset [(\Omega \times \Omega \setminus diag) \times [0,T]^2]$, denote the restriction by $\aleph^T$.
Take another base $\aleph^T_0$ in $[\Omega \times \Omega \times [0,T]^2]_{\delta,T}$, which is the tensor product
of the base  $\aleph$ with the standard Fourier base over the plane $[0,T]^2$. 
Take the decomposition of $G(x_1(t_1),x_2(t_2))$ in $\aleph^T_0$. We cut this decomposition to the segment
$k \in [0,\Delta]$ of the wave numbers with fixed upper bound $\Delta$. When (the exterior limit) $\Delta \to +\infty$, we get the total decomposition.   
By this assumption,  the function  $G(x_1(t_1),x_2(t_2))$ satisfies the analogous Condition $\odot$
(2) in $\aleph^T_0$.

Define  $m_{x_1,x_2}[\dots ]$ the integration 
of a function, which depends on $x_1,t_1,x_2,t_2 \in [\Omega \times \Omega \times [0,T]^2]_{\delta,T}$, over
all points $x_1(t_1),x_2(t_2)$ with prescribed $t_1,t_2$.

A preliminary  formula $(\ref{11})$ is the following:
\begin{eqnarray}\label{novoje2}
\begin{array}{cc}
\frac{1}{2} \chi^{[2]}_{\B} = T^{-2} \iint_{[0,+T] \times [0,+T]}  \lim_{a \to +\infty} \\ \sum_{s=0}^{\infty} \frac{(-2a)^s}{s!}  m_{x_1,x_2}\left[\frac{\partial G(x_1(t_1),x_2(t_2))}{
\partial t_1}  + \frac{\partial G(x_1(t_1),x_2(t_2))}{
\partial t_2} \right]^2 \d t_1 \d t_2. 
\end{array}
\end{eqnarray} 

The main term in the right-hand side of the formula $(\ref{novoje2})$ is 
\begin{eqnarray}\label{mainterm}
 T^{-2} \iint_{[0,+T]\times [0,+T]} m_{x_1,x_2}[G^2(x_1(t_1),x_2(t_2))] \d t_1 \d t_2. 
\end{eqnarray} 
%where the integral is taken over $[\Omega \times \Omega \times [0,T]^2]_{\delta,T}$.
The function $G^2(x_1(t_1),x_2(t_2))$ is  absolutely integrable in the largest space $[\Omega \times \Omega \times [0,T]^2]$. The  limit of  integrals over $[\Omega \times \Omega \times [0,T]^2]_{\delta,T}$, $\delta \to +0$,
convergences uniformly and absolutely. The limit of the integral
$(\ref{mainterm})$ coincides with the main term in  $(\ref{11})$, this limit does not depend on $T$. 

Consider a pair of magnetic lines $l_1, l_2$ of the length $T$, which are issued from fixed points $x_1=x_1(0)$, $x_2=x_2(0)$. 
To prove  $(\ref{novoje2})$ we assume that the Fourier spectrum of $G(t_1,t_2)$, $(t_1,t_2) \in [0,+T]^2$ is of the form:
\begin{eqnarray}\label{case}
G(t_1,t_2)=\lambda_0 + \lambda \sin(\alpha t_1 + \theta_1)\sin(\beta t_2 + \theta_2), 
\end{eqnarray}
$\quad \alpha \in \Z$; $\theta_0, \theta_1 \in [0,\pi]$ are shifts of the coordinate from the starting points $x_1(0), x_2(0)$ along the magnetic lines $l_1, l_2$.  
% We get $\lambda_0 \vert_{l_1 \cup l_2}=\lambda_{\B}(x_1,x_2)$,
%$\lambda_0(x_1,x_2), \lambda(x_1,x_2)$ are depended of positions $x_1 \in l_1, \quad x_2 \in l_2$, are depended of %magnetic lines $l_1, l_2$. 
Then the formula $(\ref{novoje2})$ for this two magnetic lines is:
$$ T^{-2}\iint [\lambda_{0}^2 + \frac{\lambda^2}{4}(\exp^2(-\infty) - \exp^2(0))] \d t_1 \d t_2. $$
The formula for the quadratic helicity is:
$$
 \iint \lambda_{\B}^2(x_1,x_2) \d \Omega \d \Omega = \frac{\chi^{[2]}_{\B}}{2},
$$
where $\lambda_{\B} = \lambda_0(x_1,x_2)$ is the mean value of the main term in $(\ref{case})$, which depends only on a pair of starting points of magnetic lines $(l_1,l_2)$.  
This proves the preliminary  formula $(\ref{novoje2})$ in a particular case. A general case assuming the Fourier base contains of only finite number of harmonics follows
from linearity, orthogonality of harmonics, and the fact that the limit $a \to +\infty$ commutes with the Fourier integral. \qed

%\begin{remark}
%A general case requires a week analog of Condition $\odot$ (2)
%for $G(t_1(x_1),x_2(t_2))$ (the same condition for $m_{x_1,x_2}[G^2(x_1(t_1),x_2(t_2))]$ is also satisfied). 
%\end{remark}

A final step of the proof of Proposition \ref{12} is a  step-by-step modification of terms in $(\ref{novoje2})$, such that this
modified formula is the restriction  of convergent integrals over the largest space $\Omega \times \Omega \times [0,T]^2$.
The convolution of the modified formula over $\Omega \times \Omega$ gives the formula  $(\ref{11})$.

Using Definition \ref{denot}, we get: $G(x_1(t_1),x_2(t_2)) =  \tilde{G}(x_1(t_1),x_2(t_2))\Psi(x_1(t_1),x_2(t_2))$.
Let us prove that the first-order derivatives $\frac{\partial G(x_1(t_1),x_2(t_2))}{\partial t_1}$,
$\frac{\partial G(x_1(t_1),x_2(t_2))}{\partial t_2}$ in the kernel of $(\ref{novoje2})$ can be replaced by
$\delta^{[2]}(\nabla_{\B_1}[\B_1\Psi(t_1,t_2)],\B_2)$, $\delta^{[2]}(\B_1,\nabla_{\B_2}[\B_2\Psi(t_1,t_2)])$, where $\B_1=\B(x_1(t_1)$, $\B_2=\B(x_2(t_2)$.

We get:
$$\frac{\partial  \hat{G}(t_1,t_2)}{\partial t_1} = \frac{\partial \left\langle \B(x_1(t_1)),\B(x_2(t_2)), x_1(t_1) - x_2(t_2)\right\rangle }{\partial t_1} = $$
$$ \left\langle\nabla_{\B_1}\B_1(t_1), \B_2(t_2), x_1 - x_2\right\rangle,$$ 
because $\frac{\d x_1}{\d t_1} = \B_1(t_1)$.  For $\frac{\partial  \hat{G}(t_1,t_2)}{\partial t_2}$  the formula is analogous.

\section{Cubic helicities} 

All helicities in this section are invariants for the group of volume-preserved diffeomorphisms of domains with magnetic fields.
There exists 3 quadratic magnetic helicities $\chi^{(2)}$ $G^4 cm^5$; $\chi^{[2]}$ $G^4 cm^2$; $\chi^2_{\B}$, $G^4 cm^8$.
Only $\chi^{(2)}$, $\chi^{[2]}$ determine second momenta (dispersions) of the asymptotic self-linking number, the square of the helicity $\chi^2_{\B}$ is the second momentum, which is the square of the first-order momentum. 
The quadratic magnetic helicity $\chi^{(2)}$, certainly, is deduced from the distribution function of Arnold's asymptotic ergodic Hopf invariant. The same time the helicity $\chi^{(2)}$ is interesting by itself, this is the $L^2$-norm of the "`Field line helicity"' (see \cite{R-Y-H-W}).

There exist $8$ different third-order momenta of the asymptotic self-linking number, which are called the cubic magnetic helicities, let us list them and indicate dimensions. 
The following diagram explains how to define the corresponding cubic helicity as a sum of corresponding
products of $3$ pairwise linking coefficients (denoted by $---$) for a collection of magnetic lines (denoted by $\odot$).

\begin{displaymath}
\chi^3_{\B}  \quad G^6 cm^{12} \quad \odot -- \odot \ \odot -- \odot \ \odot -- \odot 
\end{displaymath}

\begin{displaymath}
\chi^{(2)}\chi_{\B} \quad G^6 cm^{9} \quad \odot -- \odot -- \odot \ \odot -- \odot
\end{displaymath}

\begin{displaymath}
\begin{array}{cc}
\chi^{(3,1)}  \quad G^6 cm^{8} \quad & \odot -- \odot -- \odot\\
    &    \vert \\
    &     \odot \\
\end{array}
\end{displaymath}

\begin{displaymath}
\chi^{(3,2)} \quad G^6 cm^{8} \quad \odot -- \odot  -- \odot -- \odot  
\end{displaymath}

\begin{displaymath}
\chi^{[2]}\chi_{\B} \quad G^6 cm^{6} \quad \odot == \odot  \ \odot -- \odot  
\end{displaymath}

\begin{displaymath}
\chi^{((3,1))} \quad G^6 cm^{3} \quad \odot == \odot  -- \odot  
\end{displaymath}

\begin{displaymath}
\begin{array}{cc}
\chi^{(3,2)}  \quad G^6 cm^{3} \quad & \odot -- \odot \\
    &    \diagdown \quad \diagup \\
    &     \odot \\
\end{array}
\end{displaymath}

 \begin{displaymath}
\chi^{[3]} \quad G^6  \quad \odot \equiv\equiv \odot   
\end{displaymath}

\subsubsection*{Explanations}
$\chi^3_{\B}$ is the cube of the magnetic helicity; $\chi^{(2)}\chi_{\B}$ is the product of the quadratic magnetic helicity and the magnetic helicity; 
$\chi^{[2]}\chi_{\B}$ is the product of the quadratic momentum of magnetic helicity and the magnetic helicity; 
$\chi^{[3]}$ is the cubic momentum of the magnetic helicity, which is analogous to $\chi^{[2]}$. The difference between $\chi^{[3]}$ and $\chi^{[2]}$ is following: for $\chi^{[3]}$ the correlation tensor is
unlimited.  The only 5 cubic helicities determines  independent 3-order momenta of the asymptotic self-linking number,
the cubic helicities $\chi^3_{\B}$, $\chi^{(2)}\chi_{\B}$, $\chi^{[2]}\chi_{\B}$ are functions of quadratic helicity and helicity, they are central momenta of independent cubic helicities.
 
Let us consider an arbitrary connected graph with $n$ edges (multiple edges are admissible, edges from a vertex to itself are not admissible), graphs for $n=3$ are on the picture. The correlation tensor of a momenta of magnetic helicity is well-defined for the corresponding graph. Assume a graph satisfies the following property: 
for an arbitrary $k$ vertexes there are strongly less then $3k-3$ edges between them. In this case the correlation tensor is limited. In particular, the graph for $\chi^{[3]}$ has 2 vertexes and $3$ edges. The inequality $(2-1)3 < 3$ is not satisfied and the correlation tensor is unlimited. In the case $k=6$ consider the graph with the only edge between an arbitrary pair of vertexes. Then the number of  edges is $15$ and $3(k-1)=15$, for this graph the correlation tensor is unlimited.

\subsubsection*{Problem}
Estimate the asymptotic of independent $n$-momenta of the helicity (of the Arnold's asymptotic linking number),  and $n$-momenta for which the correlation tensor is limited, $n \to +\infty$.
\[  \]

\section{The Kolmogorov spectrum of magnetic fields}\label{Sect4}

By the Kolmogorov spectrum of magnetic fields we mean the following expression:
\begin{eqnarray}\label{Kolm}
\B(\x) = \int_{\vec{\k}} \B(\vec{\k}) \exp(\i \vec{\k} \cdot \vec{\x}) \d \vec{\k},
\end{eqnarray}
where $\vec{\k} \cdot \B(\vec{\k}) = 0$ ($\B$ is divergent-free), $\B(-\vec{\k}) = \B^{\ast}(\vec{\k})$ ($\B$ corresponds to a real solution),  $\ast$ is the complex conjugation. 
In the formula $(\ref{Kolm})$ $\B$ 
we also assume that random amplitudes $\vert\vert \B(\x) \vert\vert$  of elementary harmonics $\B(\vec{\k})$ satisfy the power low: $\vert\vert \B(\x) \vert\vert \sim k^{-\alpha}$ (for short: $_{\vec{\k}} \B \sim k^{-\alpha}$), where $\alpha$ is a real parameter.

For the turbulence with no magnetic fields the definition can be found in \cite{Fr}, the MHD turbulence is analogous (and more complicated). The goal is to explain a basic exercise "`helicity is a lower bound of magnetic energy"', Example \ref{Ex1} (compare with the Arnold inequality $(\ref{Arnoldineq})$). An analogous exercise, Example \ref{Ex2}, is defined with quadratic magnetic helicity instead of magnetic helicity.  %I do not clam that this example immediately applied %for MHD turbulence. 
A question with such a generalization was formulated by D.Sokoloff.

\begin{example}\label{Ex1}
Assume $_{\vec{\k}} \B \sim k^{-\alpha}$. Recall, $\A$ is the vector-potential for $\B$: $\rot \A = \B$. In the case $\B$ is a proper vector of the operator $\rot$ one get: $\A = k^{-1} \B$, where $k$ is the proper value of $\B$.
Then we get: $_{\vec{\k}} \A \sim k^{-\alpha-1}$. We get: $_{\vec{\k}} (\A,\B) \sim k^{2\alpha-1}$. We assume that for a fundamental domain  $vol(\Omega)=1$. We get: $_{k} \int(\A,\B) d\Omega \sim k^{-2\alpha+1}$. The helicity integral is uniformly distributed over the $k$-line in the case $\alpha=\frac{1}{2}$.
\end{example}

This example can be interpreted the following way: the distribution of the linking number of magnetic lines in $\Omega$ does not depend  on a scale when $\alpha = \frac{1}{2}$. The magnetic energy $U = \int \B^2 \d\Omega$ (dimension is $G^2 cm^3$) in this case is distributed
over the $k$-line as $\sim k^{-2\alpha + 2} = k$.  The spectrum admits an upper bound $\vert \k \vert \le \Delta$, because the
magnetic energy is finite. 
Assume that the magnetic helicity $\chi_{\B}$ is sufficiently large, then the magnetic energy has to be large, and the upper bound of the Kolmogorov spectrum has to be sufficiently large. 

To formulate a new example, let us considered Example $\ref{Ex1}$ from point of view of Gaussian kernel $G$ in  $(\ref{Gauss})$. 
With assumptions of Example \ref{Ex1} we get the following $\vec{\k}_1\times \vec{\k}_2 \times (x_1, x_2)$-distribution,  $x_1 \in \Omega$, $x_2 \in \R^3$, of the kernel $G(x_1,x_2) = (\B_1(x_1),\A_2(x_2;x_1))$:
 $$_{\vec{\k}_1 \times \vec{\k}_2 \times (x_1 x_2)} (\B_1(x_1),\A_2(x_2;x_1)) \sim k^{-2\alpha+2}.$$
Passing to the average over $(x_1,x_2)$, using $vol(\Omega)=1$, we get a distribution:   
$$_{\k_1 \times \k_2}m_{x_1,x_2}[(\B_1(x_1),\A_2(x_2;x_1))] \sim k^{-2\alpha - 1}.$$ 
Because $\k_1 = \k_2$, this gives the distribution of the helicity integral over the $k$-line: $_k \chi_{\B} \sim k^{-2\alpha + 1}$, $k=\vert \vec{\k}_1 \vert = \vert \vec{\k}_2\vert$ as above.

\begin{example}\label{Ex2}

 A distribution of the integral kernel $G^2(x_1,x_2)$ of the main term in the formula $(\ref{11})$ is well-defined over the Cartesian product $\vec{\k}_1 \times \vec{\k}_2 \times \vec{\k}'_1 \times \vec{\k}'_2$. 
 Proper vectors give contribution to $G^2(x_1,x_2)$ only with its square, this gives  $\vec{\k}_1 \times \vec{\k}_2$-distribution.
With assumptions of Example \ref{Ex1} we get the following distribution of the kernel $$G^2(x_1,x_2)=(\B_1(x_1),\A_2(x_2;x_1))$$ 
at a prescribed point $(x_1,x_2) \in \Omega \times \R^3$:  
$$_{\vec{\k}_1 \times \vec{\k}_2  \times (x_1,x_2)} G^2(x_1,x_2) \sim k^{-4\alpha + 4}.$$ 
After the average of the distribution over $(x_1,x_2)$ we get the following $\vec{\k}_1 \times \vec{\k}_2 $-distribution:
$$_{\vec{\k}_1 \times \vec{\k}_2}\delta^{[2]}(\B_1,\B_2) \sim k^{-4\alpha + 1}.$$ This describes the distribution 
$$ _{k} \delta^{[2]}(\B_1,\B_2) \sim k^{-4\alpha + 6}$$
of the main term in $(\ref{11})$, where $k$ is the module of the vector $\vec{\k}_1 \times \vec{\k}_2 $.

Let us describe the distribution $(\ref{11})$. 
We have to take two collections of random vectors $\{\B_{\vec{\k}_{1,a}};\B_{\vec{\k}_{1,1}}, \dots, \B_{\vec{\k}_{1,i}}\}$
$\{\B_{\vec{\k}_{2,a}},\B_{\vec{\k}_{2,1}}, \dots, \B_{\vec{\k}_{2,j}}\}$ in the Kolmogorov spectrum. This collections determine  random
distribution of the term
$\delta^{[2]}(\B^{(i)}_1,\B^{(j)}_2)$, $i+j=s$ in the right-hand side of the formula $(\ref{11})$.
$$ \B_1^{(i)}(\vec{\k}_{1,a},\vec{\k}_{1,1}, \dots, \vec{\k}_{1,i}) = \nabla_{\B_{\vec{\k}_{1,i}}} \dots \nabla_{\B_{\vec{\k}_{1,1}}} \B_{\vec{\k}_{1,a}}, $$
$$ \B_2^{(j)}(\vec{\k}_{2,a},\vec{\k}_{2,1}, \dots, \vec{\k}_{2,j}) = \nabla_{\B_{\vec{\k}_{2,j}}} \dots \nabla_{\B_{\vec{\k}_{2,1}}} \B_{\vec{\k}_{2,a}}. $$
In the case $i=0$, $j=0$ we get the distribution above for $\B_1=\B_{1,a}$, $\B_2=\B_{2,a}$.
%$$ \B_2(\vec{\k}_b,\vec{\k}_1, \dots, \vec{\k}_i) = \nabla_{\B_{\vec{\k}_i}} \dots \nabla_{\B_{\vec{\k}_1}}  %\B_{\vec{\k}_b}, $$
%$$ \B'_2(\vec{\k}'_b,\vec{\k}'_1, \dots, \vec{\k}'_j) = \nabla_{\B_{\vec{\k}'_j}} \dots \nabla_{\B_{\vec{\k}'_1}}  %\B_{\vec{\k}'_b}. $$

For $s=1$ (we consider one of the two similar distributions with $i=1$, $j=0$) we get:
$$_{k} \delta^{[2]}(\B^{(1)}_1,\B_2) \sim _{k}\delta^{[2]}(\B_{1,a},\B_{2,a}) \overline{\sin^2(\theta)}C,$$
where $\theta$ is a random angle on the unit sphere between vectors $\B_{1,1}$ and $\B_{1,a}$, a positive constant $C$ is dimensionless, $\overline{g(\theta)}$ is the mean value of a distribution $g(\theta)$.
%The derivatives $\nabla_{\B_{.,.}}$ in expression $(\ref{11})$ for elementary harmonics are simplified. 

 By induction for $s \ge 2$ (we assume that $i \ge 1$ to get an inductive step $i-1 \mapsto i$) we get:
\begin{eqnarray}\label{19}
\begin{array}{ccc}
_{k} \delta^{[2]}(\B^{(i)}_1,\B^{(j)}_2) \sim \\
\\
 _{k} \delta^{[2]}(\B_1,\B_2)\overline{\vert\sin(\theta_{1,1})\vert}^2 \dots \overline{\vert\sin(\theta_{1,i-1})\vert}^2 \quad \overline{\sin^2(\theta_{1,i})} \quad \overline{\vert\sin(\theta_{2,1})\vert}^2 \dots \overline{\vert \sin(\theta_{2,j})\vert}^2 2^{s-1}C^{s},
\end{array}
\end{eqnarray}  
where  $\theta_{1,1}, \dots \theta_{1,i},\theta_{2,i}, \dots,\theta_{2,j}$ are latitudes on the coordinate unit sphere, pointed by the vectors $\B_{1,1}, \dots, \B_{1,i}$ with $\B_{1,a}$, and by $\B_{2,1}, \dots, \B_{2,j}$ with $\B_{2,a}$. The angles have a common distribution and the convolution of $(\ref{19})$ is distributed as 
$$ \sim _{k}\delta^{[2]}(\B_1,\B_2) \quad \frac{\overline{\sin^2(\theta)}}  {2(\overline{\vert\sin(\theta)\vert})^2} \- C^s.$$
The value of the expression $(\ref{11})$ is distributed as $\frac{1}{3}$ of  the main term (this follows from the formulas: $\int_{S^2} \sin^2(\theta) \d S^2  = \frac{1}{3} \int_{S^2} \d S^2$;
$\int_{S^2} \vert \sin(\theta) \vert \d S^2 = \frac{1}{2} \int_{S^2} \d S^2$). After we pass to the $k$-line,  terms in $(\ref{11})$ are distributed as the main term by the formula: 
\begin{eqnarray}\label{asimpt}
_{k} \chi^{[2]}_{\B} \sim k^{-4\alpha +6}.
\end{eqnarray}
 The uniformly distribution for $\chi^{[2]}_{\B}$ is in the case $\alpha = \frac{3}{2}$.

%Assume the magnetic energy is distributed over the $k$-line as $\sim k^{-\frac{1}{2}}$, the magnetic energy
%is distributed as $\sim k$ as in Example \ref{Ex1}.   

An elementary magnetic vector in $(\ref{Kolm})$ admits the complex and the real component. As the result, for a given $\vec{\k}$ we get two magnetic harmonics with positive (right) and negative (left) helicity. Assume that the contribution of left and right harmonics for all $\vec{\k}$ in  $(\ref{Kolm})$ is opposite. Then the Example $\ref{Ex1}$ gives us no estimate  of the magnetic energy from above, because $\chi_{\B}=0$. The quadratic helicity is an invariant for ideal MHD, assume its value is sufficiently large. In this case the lower bound of the spectra can be estimated.

Cut-out wave vectors with $\Delta' \vert\vert k \vert\vert < \Delta$, $\Delta >> 1$, $0 < \Delta' <<1$. The interior limit $a \to +\infty$, $a >> \Delta$, and the exterior limit $\Delta \to +\infty$, $\Delta' \to 0+$ are defined the 2-variables limit.

\end{example}

%\section{Quadratic helisity as a function on the Kolmogorov's spectrum}

%\section{Conclusion and discussion}

\end{document}